\documentstyle[twocolumn,aps,epsfig,prl,amsfonts,floats]{revtex}

\newcommand{\av}[1]{\left\langle{#1}\right\rangle}
\newcommand{\binomial}[2]{{{#1} \choose {#2}}}
\newcommand{\pref}[1]{(\ref{#1})}

\begin{document}

\title{Random Costs in Combinatorial Optimization} \draft
\author{Stephan~Mertens\thanks{ http://itp.nat.uni-magdeburg.de/\~{}mertens,
    email:Stephan.Mertens@Physik.Uni-Magdeburg.DE}}

\address{Universit\"at Magdeburg, Inst.\ f.\ Theoretische Physik,
  Universit\"atsplatz 2, D-39106 Magdeburg, Germany }

\date{\today}

\maketitle

\begin{abstract}
  The random cost problem is the problem of finding
  the minimum in an
  exponentially long list of random numbers. By definition, this
  problem cannot be solved faster than by exhaustive
  search. It is shown that a classical NP-hard optimization problem, 
  number partitioning, is essentially equivalent
  to the random cost problem.  This explains the bad performance of
  heuristic approaches to the number partitioning problem and allows
  us to calculate the probability distributions of the optimum and
  sub-optimum costs.
\end{abstract}


Recent years have witnessed an increasing interaction among the disciplines of
discrete mathematics, computer science and statistical physics. Particularly the
methods and concepts developed in spin glass theory have been applied
successfully to problems from combinatorial optimization \cite{mezard:etal:87}.
An optimization problem is defined by a set ${\cal X}$ (the domain) of feasible
solutions $\sigma\in {\cal X}$ and a real valued function $H$ on $X$. The
minimization problem then is: Find that $\sigma\in {\cal X}$ which minimizes
$H(\sigma)$. In combinatorial optimization ${\cal X}$ is always countable. For
minimization problems, $H$ is called the cost-function, physicists call it
Hamiltonian or energy.

Most of the problems in combinatorial opimization are NP-hard
\cite{papadimitriou:steiglitz:82}, which means that no algorithm is
known that solves the problem significantly faster than exhaustive
search of the domain.  Although it has not been proven, it is widely
believed that for NP-hard problems faster algorithms do not exist
\cite{garey:johnson:79,papadimitriou:94}.  A problem for which this
can be proven is the {\it random cost problem}: Here the cost function
is a table of random numbers and the role of $\sigma$ is reduced to an
index. It can be proven, that one has to look at every number in the
table to find the minimum \cite{clr:algorithms}. Furthermore it is
obvious that there is no better heuristic than repeated random lookup.
In this sense the random cost problem is harder than many other
NP-hard problems, for which much better heuristics do exist.

In this contribution it is shown that the random cost problem is not an
artificial toy problem but a valid description of at least one classical problem
from combinatorial optimization: the number partitioning problem, NPP.

Number partitioning is one of Garey and Johnson's \cite{garey:johnson:79} six
basic NP-complete problems that lie at the heart of the theory of
NP-completeness. It is defined as follows: Given a set $\{a_1,a_2,\ldots,a_N\}$
of positive numbers, find a partition, i.e.\ two disjoint subsets ${\cal A}_1$
and ${\cal A}_2$ such that the residue
\begin{equation}
  \label{eq:costfunction}
  E = \Big|\sum_{a_j\in {\cal A}_1} a_j - \sum_{a_j\in {\cal A}_2} a_j\Big|
\end{equation}
is minimized. In the balanced number partioning problem, the optimization is
restricted to partitions with $|{\cal A}_1|=|{\cal A}_2|=N/2$ ($N$ even).  A
partition can be encoded by Ising spins $s_j=\pm 1$: $s_j=1$ if $a_j\in {\cal
  A}_1$, $s_j=-1$ otherwise. The cost function then reads
\begin{equation}
  \label{eq:costfunction_spins}
  E = \Big|\sum_{j=1}^Na_js_j\Big|,
\end{equation}
and the minimum partition is equivalent to the ground state of the Hamiltonian
\begin{equation}
  \label{eq:hamiltonian}
  H = E^2 = \sum_{i,j=1}^N s_i\,a_ia_j\,s_j.
\end{equation}
In statistical mechanics, this is an infinite range Ising spin glass with Mattis-like, antiferromagnetic
couplings $J_{ij}=-a_ia_j$ \cite{fu:89,ferreira:fontanari:98,ferreira:fontanari:99,mertens:98a}.

The computational complexity of the NPP depends on the number of bits needed to
encode the numbers $a_j$.  Numerical simulations show, that for independent,
identically distributed (i.i.d.) random $b$-bit numbers $a_j$, the solution time
grows exponentially with $N$ for $N < b$ (roughly) and polynomially for $N > b$
\cite{gent:walsh:96,korf:95,korf:98}. The transition from the ``hard'' to the
computational ``easy'' phase has some features of a phase transition in physical
systems.  Phase transitions of this kind have been observed in numerous
NP-complete problems \cite{cheeseman:etal:91,gent:walsh:95,monasson:etal:99},
and can often be analyzed quantitatively in the framework of statistical
mechanics. Compared to other problems, this analysis is surprisingly simple for
the number partitioning problem \cite{mertens:98a}.

Here we concentrate on the computationally hard regime $N \ll b$, i.e.\ we
consider the $a_j$ to be real numbers of ``infinite'' precision. For this case,
Karmarkar et al.~\cite{karmarkar:etal:86} have proven that the median value of
the optimum residue $E_1$ is ${\cal O}(\sqrt{N}\cdot 2^{-N})$ for the
unconstrained and ${\cal O}(N\cdot 2^{-N})$ for the balanced partitioning
problem. Their proof yields no results on the distribution of $E_1$, however, or
at least on its average value.  Numerical simulations
\cite{ferreira:fontanari:98} indicate, that the relative width of the
distribution of $E_1$, defined as
\begin{equation}
  \label{eq:def-r}
  r := \frac{\sqrt{\av{E_1^2} - \av{E_1}^2}}{\av{E_1}}
\end{equation}
where $\av{\cdot}$ denotes the average over the $a_j$'s, tends to a finite value
in the large $N$ limit, more precisely: $\lim_{N\to\infty} r = 1$, for both the
unconstrained and the balanced partitioning problem. This means, that the ground
state energy is a non self averaging quantity.

Another surprising feature of the NPP is the bad performance of heuristic
algorithms \cite{johnson:etal:91,ruml:etal:96}.  The best known heuristic, the
differencing method \cite{karmarkar:karp:82,korf:98}, yields partitions with
expected residue ${\cal O}(N^{-a\log N})$, $a>0$ for $a_j$ distributed uniformly
between $0$ and $1$. This is still bad compared to ${\cal O}(\sqrt{N}\cdot
2^{-N})$ for the true optimum.

In this contribution we show that all these features can be understood
qualitatively and quantititavely by the observation, that number partitioning is
essentially equivalent to a random cost problem.  Our line of reasoning closely
follows Derrida \cite{derrida:81,gross:mezard:84}, who introduces the random
energy model (REM) in spin glass theory. The random cost problem is the
optimization counterpart of the REM, with some modifications, as we will see
below.


In the balanced NPP, the energies are distributed according to
\begin{equation}
  \label{eq:def-rho}
  P(E) = {\binomial{N}{N/2}}^{-1}{\sum_{\{s_j\}}}' 
  \av{\delta(E-|\sum_ja_js_j|)},
\end{equation}
where the primed sum runs over all spin configurations with $\sum s_j = 0$.  The
symmetry of the problem and our assumption of i.i.d.\ random variables $a_j$
allow us to write
\begin{equation}
   \label{eq:sym-rho}
  P(E) = 2\cdot \av{\delta\big(E-\sum_{j=1}^{N/2}(a_j-a_{N/2+j})\big)} \cdot 
  \Theta(E),
\end{equation}
where $\Theta$ denotes the step-function, $\Theta(x) = 1$ for $x\geq0$ and
$\Theta(x) = 0$ for $x<0$.  The symmetrization $a_{\mathrm sym}$ of $a$, i.e.\ 
the random variable distributed as the result of subtracting two independent
variables $a_1-a_2$, has mean $0$ and variance $2\sigma^2$ with
$\sigma^2=\av{a^2}-\av{a}^2$.  If $g_k$ denotes the density of the $k$-th
partial sum of $a_{\mathrm sym}$ we can write $P(E) = 2g_{N/2}(E)\Theta(E)$,
which according to the central limit theorem becomes
\begin{equation}
  \label{eq:P-balanced}
  P(E) = \frac{2}{\sqrt{2\pi\sigma^2 N}}\exp\Big(-\frac{E^2}{2\sigma^2 N}\Big)
\cdot\Theta(E) + {\cal O}(N^{-3/2})
\end{equation}
for large values of $N$. The energies in the unconstrained NPP follow the same
distribution but with $\sigma^2$ replaced by $\av{a^2}$.


The probability density of finding energies $E_1$ and $E_2$ is
\begin{eqnarray}
  \label{eq:def-rho-E1-E2}
P(E_1,E_2) &=& 4\cdot\Theta(E)\Theta(E') {\binomial{N}{N/2}}^{-2} \cdot\\
 \nonumber&\cdot&
  {\sum_{\{s_j\}}}'{\sum_{\{s'_j\}}}' \av{\delta(E_1-\sum_ja_js_j)
   \cdot \delta(E_2-\sum_ja_js'_j)}
\end{eqnarray}
for the balanced NPP.  Again we use the gauge invariance to state that each term
in the above sum depends on $\{s_j\}$ and $\{s'_j\}$ only through the overlap
\begin{equation}
  \label{eq:def_q}
  Q = \sum_{j=1}^{N}s_js_j'.
\end{equation}
Then
\begin{equation}
  \label{eq:P-EE-sum}
  P(E_1,E_2) = \frac{4\Theta(E_1)\Theta(E_2)}{\binomial{N}{N/2}}{\sum_{Q=-N}^{N}}' 
  {\binomial{N/2}{\frac{N+Q}{4}}}^2 
P_Q(E_1, E_2),
\end{equation}
where the primed sum denotes summation over $Q=-N,-N+4,\ldots,N-4,N$ and
\begin{equation}
  \label{eq:PQexact}
  P_Q(E_1, E_2) = \frac{1}{2}\cdot g_{(N+Q)/4}(\frac{E_1+E_2}2) \cdot g_{(N-Q)/4}
  (\frac{E_1-E_2}2)
\end{equation}
(see above for a definition of $g$). For large values of $N$, the central limit
theorem tells us that
\begin{equation}
  \label{eq:PQclt}
  P_Q(E_1, E_2) = \frac{1}{2\pi\sigma^2 N \sqrt{1-q^2}}\cdot e^{-\frac{E_1^2+E_2^2-2E_1E_2q}
{2\sigma^2N(1-q^2)}}
\end{equation}
with $q=Q/N$. In the same limit we may apply Stirling's formula to the binomial
coefficients and replace the sum over $Q$ by an integral over $q$:
\begin{eqnarray}
  \label{eq:PEEasympt}
  P(E_1,E_2) &=& \frac{2\Theta(E_1)\Theta(E_2)}{\pi^2\sigma^2N}\sqrt{\frac{\pi N}2} \cdot 
\\\nonumber
          & & \cdot \int \frac{dq}{1-q^2} \, e^{-\frac{E_1^2+E_2^2-2E_1E_2q}{2\sigma^2
N(1-q^2)}} \, e^{-N f(q)}
\end{eqnarray}
with
\begin{equation}
  \label{eq:def-f}
  f(q) = \frac12 (1+q)\ln(1+q) + \frac12 (1-q)\ln(1-q).
\end{equation}
The integral can be evaluated using the saddle point approximation.  For $E_1$
and $E_2$ both ${\cal O}(\sqrt{N})$, the saddlepoint is at $q=0$:
\begin{equation}
  \label{eq:PEEsaddle}
  P(E_1,E_2) = \frac{2\Theta(E_1)\Theta(E_2)}{\pi\sigma^2 N} \exp{\Big(-\frac{E_1^2+E_2^2}
{2\sigma^2N}\Big)}
\end{equation}
i.e.\ $P(E_1,E_2)=P(E_1)\cdot P(E_2)$. A similar calculation shows that
$P(E_1,E_2)$ factorizes for the unconstrained NPP, too. Note that for $E = {\cal
  O}(N)$ the saddle point is no longer at $q=0$ and $P(E_1,E_2)$ does not
factorize. This is plausible, since energies ${\cal O}(N)$ can only be achieved
by putting a number ${\cal O}(N)$ of the lowest values $a_j$ into one partition
and a number ${\cal O}(N)$ of the largest values in the complement. The
corresponding spin sequences then have an overlap ${\cal O}(N)$.

The two basic properties that lead to the factorization of $P(E,E')$ are the
gauge invariance, i.e.\ the fact that $\av{\delta(E-\sum a_js_j)\delta(E'-\sum
  a_js'_j)}$ only depends on the overlap $q$ of the sequences, and the entropic
dominance of the $q=0$ contributions.  Both properties persist if one considers
the probability distributions of three or more levels, so we claim that
$P(E_1,E_2,\ldots,E_k)$ factorizes as well. Instead of providing a formal
derivation, we consider this as an assumption and discuss its consequences.


Motivated by the factorization of the distribution of energies, we may now
specify our random cost problem: Given are $M={\cal O}(2^N)$ random numbers
$E_i$, {\it independently} drawn from the density $P(E)$,
Eq.~\pref{eq:P-balanced}.  Find the minimum of these numbers.  The
correspondence to the NPP requires $M=\frac12\binomial{N}{N/2}$ for the balanced
and $M=2^{N-1}$ for the unconstrained case.

\begin{figure}[thb]
  \includegraphics[width=\columnwidth]{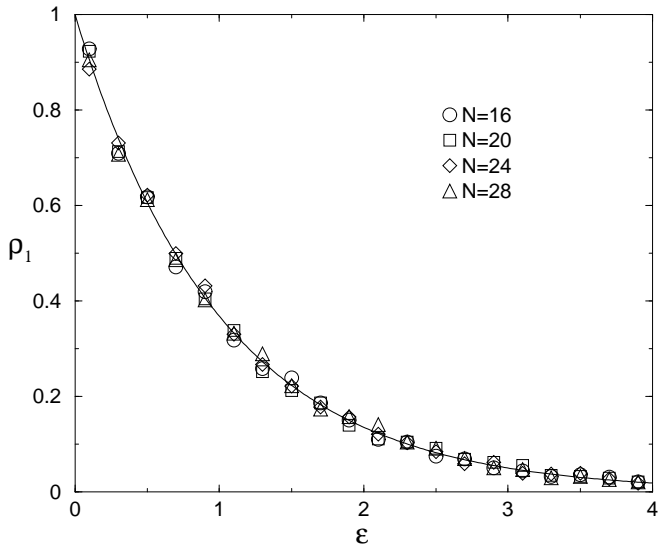}
\caption{
  Distribution of scaled ground state energies for the balanced number
  partioning problem. The solid line is given by Eq.~\pref{eq:rho1}, the symbols
  are averages over $10^4$ random samples.
\label{fig:rho1}}
\end{figure}

Let $E_k$ denote the $k$-th lowest energy of an instance of our random cost
problem.  The independence of the $E_i$ enables us to write
\begin{equation}
  \label{eq:basic-rho1}
  \rho_1(E_1) = M\cdot P(E_1)\cdot \Big(1-\int_0^{E_1} P(E') dE'\Big)^{M-1}
\end{equation}
for the probability density $\rho_1$ of the minimum energy.  $E_1$ must be small
to get a finite r.h.s.\ in the large $M$ limit. Hence we may write
\begin{eqnarray*}
  \rho_1(E_1) &\approx& M\cdot P(0) \cdot \Big(1-E_1 P(0)\Big)^{M-1} \\
              &\approx& M\cdot P(0) \cdot e^{-M P(0) E_1}.
\end{eqnarray*}
This means that the probability density of the scaled minimal energy,
\begin{equation}
  \label{eq:def-epsilon}
  \varepsilon_1 = M \cdot P(0) \cdot E_1
\end{equation}
for large $M$ converges to a simple exponential distribution,
\begin{equation}
  \label{eq:rho1}
  \rho_1(\varepsilon) = e^{-\varepsilon} \cdot \Theta(\varepsilon).
\end{equation}
Note that a rigorous derivation from Eq.~(\ref{eq:basic-rho1}) to
Eq.~(\ref{eq:rho1}) can be found in any textbook on extreme order statistics
\cite{galambos:book}.  Along similar lines one can show that the density
$\rho_k$ of the $k$-th lowest scaled energy is
\begin{equation}
  \label{eq:rhok}
  \rho_k(\varepsilon) = \frac{\varepsilon^{k-1}}{(k-1)!} \cdot e^{-\varepsilon} \cdot 
  \Theta(\varepsilon)
  \qquad k = 2,3,\ldots.
\end{equation}

Let us compare Eqs.~\pref{eq:rho1} and \pref{eq:rhok} with other analytical and
numerical results. From the moments of the exponential distribution
Eq.~\pref{eq:rho1}, $\av{\varepsilon^n}=n!$, we get
\begin{equation}
  \label{eq:2nd-moment}
  r = \frac{\sqrt{\av{E_1^2} - \av{E_1}^2}}{\av{E_1}} = 1,
\end{equation}
in perfect agreement with the numerical findings of Ferreira and Fontanari
\cite{ferreira:fontanari:98}. The average ground state energy is
$\av{E_1}=1/(M\cdot P(0))$, which gives
\begin{equation}
  \label{eq:avE1-balanced}
  \av{E_1} = \pi\cdot\sigma\cdot N \cdot 2^{-N}
\end{equation}
for the balanced and
\begin{equation}
  \label{eq:avE1-unconstrained}
  \av{E_1} =  \sqrt{2\pi\av{a^2}}\cdot\sqrt{N}\cdot 2^{-N}
\end{equation}
for the unconstrained NPP. Again this is in very good agreement with numerical
\cite{ferreira:fontanari:98} and analytical \cite{mertens:98a} results.

\begin{figure}[thb]
  \includegraphics[width=\columnwidth]{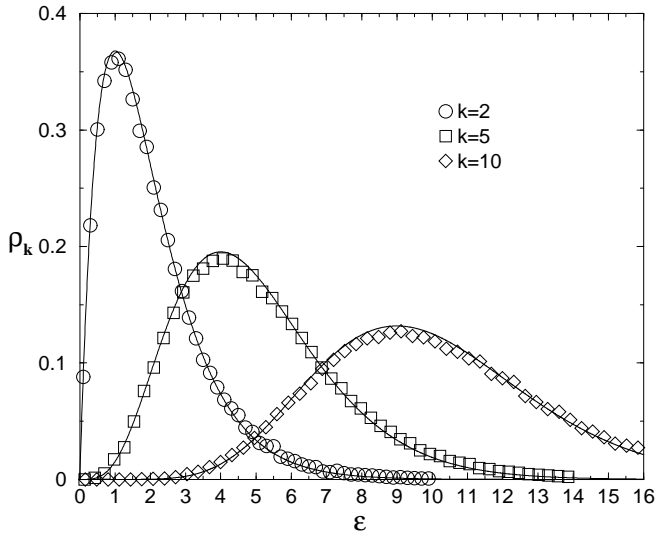}
\caption{
  Distribution of scaled $k$-th lowest energy for the balanced number partioning
  problem. The solid lines are given by Eq.~\pref{eq:rhok}, the symbols are
  averages over $10^5$ random samples of size $N=24$.
\label{fig:rhok}}
\end{figure}

To check that the random cost ansatz does not only give the correct first and
second moment of $E_1$, we calculated the distribution of $E_1$ and higher
energies numerically. Figs.~\pref{fig:rho1} and \pref{fig:rhok} display the
results for the balanced NPP.  Equivalent plots for the unconstrained NPP look
similar.  The agreement between the numerical data and Eqs.~\pref{eq:rho1} and
\pref{eq:rhok} is convincing. The algorithm used to solve larger instances of
the balanced NPP is described in \cite{mertens:99a}.

All in all, the random cost problem seems to be a valid alternative formulation
of the number partitioning problem. This correspondence not only provides new
analytic results on the NPP but also has some consequences for the dynamics of
algorithms: Any heuristic that exploits a fraction of the domain, generating and
evaluating a series of feasible configurations, cannot be better than random
search.  The best solution found by random search is distributed according to
Eq.~(\ref{eq:basic-rho1}), i.e.\ the average heuristic solution should approach
the true optimum no faster than ${\cal O}(1/M)$, $M$ being the number of configurations
generated.  Note that the best known heuristic,
the complete Karmarkar-Karp differencing \cite{korf:98,mertens:99a} converges slower,
namely like ${\cal O}(1/M^\alpha)$ with $\alpha < 1$ to the true optimum. It would 
be interesting to check whether simple random search really converges faster.
Beyond number partitioning, the dynamics of heuristic
algorithms for other combinatorial optimization problems may be considered as a
signature of a corresponding random cost problem, possibly with a differing single cost
distribution.

With its focus on costs rather than configurations, our random cost problem is
very similar to Derrida's random energy model from statistical mechanics
\cite{derrida:81,gross:mezard:84}, with an important difference: the single
energy distribution in Derrida's model is Gaussian, i.e.\ in principle it allows
arbitrary low energies. The random cost formulation of the NPP on the other hand
leads to a strict lower bound for the energies. As a consequence, both models
belong to different universality classes with respect to their asymptotic order
statistics \cite{galambos:book}. The replica-method from statistical mechanics
solves the Gaussian random energy model but fails for bounded distributions like
the one encountered here \cite{bouchaud:mezard:97}. It is an open problem how to
modify the replica method in order to reproduce the statistical mechanics of the
number partitioning problem \cite{mertens:98a}.

\acknowledgements The author appreciates stimulating discussions with Andreas
Engel and Olivier Martin. The manuscript considerably benefits from the referee
reports and from a comment by Jean-Philippe Bouchaud and Marc M\'ezard.

\bibliographystyle{prsty} \bibliography{complexity,cs}

\end{document}